\documentclass[namedreferences]{solarphysics}
\usepackage[optionalrh,solaromanenum]{spr-sola-addons} 
\usepackage{graphicx}                    

\newcommand{\dd}{{\mathrm d}}

\begin{document}

\begin{article}

\begin{opening}

\title{Persistent Near-Surface Flow Structures from Local Helioseismology}

%

\author[addressref={1},corref,email={rhowe@nso.edu}]{\inits{R.}\fnm{Rachel }\lnm{Howe}}
\author[addressref={2}]{\inits{R.W.}\fnm{R.W.}~\lnm{Komm}}
\author[addressref={3}]{\inits{D.}\fnm{D.~}~\lnm{Baker}}
\author[addressref={3}]{\inits{L.}\fnm{L.~}~\lnm{Harra}}
\author[addressref={3,4,5}]{\inits{L.}\fnm{L.~}~\lnm{van~Driel-Gesztelyi}}
\author[addressref={6}]{\inits{R.S.}\fnm{R.S.~}~\lnm{Bogart}}
%
\runningauthor{R. Howe {\it et al.}}
\runningtitle{Persistent Flow Structures from Ring Diagrams}

\address[id={1}]{School of Physics and Astronomy, University of Birmingham, Edgbaston, Birmingham B15 2TT, United Kingdom}
\address[id={2}]{National Solar Observatory, 950 N. Cherry Avenue, Tucson, AZ 85719, USA}
\address[id={3}]{UCL -- Mullard Space Science Laboratory,
Holmbury St Mary, Dorking, Surrey, RH5 6NT, UK}
\address[id={4}]{Observatoire de Paris, LESIA, CNRS, UPMC Univ. Paris 06, Univ. Paris-Diderot, Meudon, France}
\address[id={5}]{Konkoly Observatory, Hungarian Academy of Sciences, Budapest, Hungary}
\address[id={6}]{HEPL Solar Physics, 452 Lomita Mall, Stanford University, Stanford, CA 94305-4085, USA}

%

\begin{abstract}
Near-surface flows measured by the ring-diagram technique of local helioseismology show structures that persist over multiple rotations. We examine these
phenomena using data from the {\em Global Oscillation Network Group} (GONG) and
the {\em Helioseismic and Magnetic Imager} (HMI) and show that a correlation analysis of the structures can be used to estimate the rotation rate as a function of latitude, giving a result consistent with the near-surface rate from global helioseismology and slightly slower than that obtained from a similar analysis of the surface magnetic field strength. At latitudes of 60$^{\circ}$ and above the HMI flow data reveal a strong signature of a two-sided zonal flow structure. 
This signature may be related to recent reports of ``giant cells'' in solar convection.
\end{abstract}

%

\end{opening}

%

\section{Introduction}
Horizontal flows in the solar envelope are seen at a wide range of scales, from the global scale of differential rotation and poleward meridional flow down to the granular level. Many of these flows are modulated by surface activity and/or the solar cycle; at medium scales this manifests in the so-called torsional oscilation pattern of zonal flows (\opencite{1980ApJ...239L..33H}; for a review of the helioseismic observations see \inlinecite{2009LRSP....6....1H} and references therein) and the solar-cycle modulation of the meridional flow (\opencite{2004ApJ...603..776Z}; \opencite{2006ApJ...638..576G}; \citeyear{2008SoPh..252..235G}; \opencite{2010Sci...327.1350H}; \opencite{2010ApJ...717..488B}; \opencite{2013SoPh..287...85K}), while at smaller scales we see flows associated with active regions \cite[{\textit{e.g.}}][]{2004SoPh..220..371H}. There have also been reports of flow patterns attributed to ``giant cells'' in the subsurface convection, most recently for example by \inlinecite{2013Sci...342.1217H}. The aim of the current study was to look for relationships between the near-surface horizontal flows detected by helioseismology, surface magnetic activity, and coronal activity as they evolve over a number of years. In particular, we are interested in features that persist over multiple rotations. Such features, unless they are rotating rigidly at the Carrington rate, will appear to drift in longitude from one Carrington rotation to the next; this drift therefore allows us to estimate the associated rotation rate. 

\section{Data}

\subsection{Helioseismic Data}
The ring-diagram technique of local helioseismology \citep{1988ApJ...333..996H}
uses the 3D oscillation spectrum of small areas of the solar surface, usually tracked across the disk at a speed close to the surface rotation rate, to 
estimate local flows and structure. The technique is so called because the spectrum shows concentric ``rings'' of power (one for each radial order [$n$]) when sliced at constant frequency. Flows displace these rings and structural changes cause them to change size. The results of fits at different radial orders and spatial wavelengths can then be inverted to produce estimates of the horizontal flows in the analysis region as a function of depth.

The ``dense-pack'' approach, where the solar disk is covered by overlapping 15$^{\circ}$ patches observed for approximately one day, was pioneered by \inlinecite{2002ApJ...570..855H} and has become the standard for pipeline analysis. The inversion procedure returns results at 16 target depths ranging from 0.63 to 15.79\,Mm. In this work we use ring-diagram results from the {\em Global Oscillation Network Group} (GONG) and the {\em Helioseismic and Magnetic Imager} (HMI:~\opencite{2012SoPh..275..229S}) onboard the {\em Solar Dynamics Observatory}.

\subsubsection{GONG}
GONG has been producing continuous
ground-based network data for local helioseismology since mid-2001, with $1024\times 1024$-pixel detectors capable of detecting oscillations up to about $\ell=1000$. The analysis tracks 189 patches of $16^{\circ}\times 16^{\circ}$, centered at locations 
$\pm 52.5^{\circ}$ from the Equator and the central meridian and spaced at $7.5^\circ$ in heliographic latitude and longitude, at the Snodgrass rotation rate for each 1664-minute ``ring day.'' The patches are apodized to circles of diameter 
$15^{\circ}$.

\subsubsection{HMI}
HMI has been in operation since April 2010, collecting magnetic and Doppler velocity data at $4096\times4096$-pixel resolution. The pipeline processing for ring diagrams has been described by \inlinecite
{ASNA:ASNA200610741} and \citeauthor{2011JPhCS.271a2008B} (\citeyear{2011JPhCS.271a2008B}, \citeyear{2011JPhCS.271a2009B}). This processing differs from the more traditional dense-pack approach in that the patches are tracked at the Carrington rather than the Snodgrass rotation rate -- that is, the tracking rate does not depend on the latitude. Also, because of the higher resolution of HMI the analysis can be extended to higher latitudes, with patches at up to $\pm 75^{\circ}$ available for part of the year. The longitude grid used at higher latitudes differs from that at low latitude and changes throughout the year.

\subsection{Magnetic Data}
The magnetic data used here are taken from the magnetic synoptic charts from the {\em Michelson Doppler Imager} (MDI:~\opencite{1995SoPh..162..129S}) up to the end of its operation in 2011  and from HMI thereafter. The signed longitudinal magnetic field strength is  averaged over heliographic latitude and Carrington longitude ranges corresponding to the (nominal) locations of each ring-diagram synoptic-map patch in heliographic co-ordinates.

\section{Analysis and Results}

\subsection{Longitude--Time Plots}

For this analysis the ring-diagram zonal and meridional flow data were first combined into 
synoptic charts.
The HMI data at higher latitudes were interpolated to the same longitude grid as the lower-latitude and GONG data. 
We next formed
a weighted mean velocity [$\bar V(d,T,\theta)$], where $V$ stands for the
zonal or meridional velocity estimate from the inversion
 over all of the 
$n_\phi$ longitude steps [$\phi$] for each 
depth [$d$], 
Carrington rotation [$T$], 
and latitude [$\theta$];
\begin{equation}
{<{V}(d,T,\theta)>}={{{\sum_{i=1}^{n_\phi}}{{V(d,T,\theta,\phi_i)}\over{\sigma_v^2(d,T,\theta,\phi_i)}}}\over{\sum_{i=1}^{n_\phi}{1\over{\sigma^2_V(d,t,\theta,\phi_i)}}}},
\end{equation}
where $\sigma_V$ is the standard deviation of $V$. We then subtracted this from the velocity values to obtain a residual
\begin{equation}
\delta V(d,T,\theta,\phi)=V(d,T,\theta,\phi)-<{V}(d,T,\theta)>. 
\end{equation}
The purpose of this step was to remove the first-order effects of latitudinal variation in the flows, annually-varying projection effects \citep{2006SoPh..236..227Z,2015SoPh..290.1081K}, and solar-cycle-related 
changes in the overall flow patterns, while revealing the disruptions of the flows around local features.
Finally the zonal and meridional velocity residuals were averaged over all of the $n_d$ depth locations [$d$] available from the inversions to obtain 
an averaged residual 
\begin{equation}
{<\delta V(T,\theta,\phi)>}={\sum_{i=1}^{n_d}{\delta V(d_i,T,\theta)\over{\sigma^2_{\delta V}(d_i,T,\theta)}}\over
{\sum_{i=1}^{n_d}{\sigma^2_{\delta V}(d_i,T,\theta)}}}.
\end{equation}
The $1/\sigma^2$ weighting used in the averages results in a heavier weighting for the shallower depths where the uncertainties are smaller.

\subsubsection{GONG}

The GONG ring data cover the period from Carrington Rotation (C.R.) 1979 in August 2002, just after the maximum of Solar Cycle 23, to C.R. 2139 in 2013, around the considerably weaker maximum of Cycle 24. In between the two maxima falls the unusually quiet and prolonged solar minimum of 2008\,--\,2009.

In order to 
follow the migration of structures in longitude over this period, 
we plot the residual velocity as a function of Carrington longitude and time for each latitude. In such plots, a location rotating at a rate [$\Omega$] that is different from the Carrington rate [$\Omega_{\mathrm{CR}}$] will drift in Carrington longitude at a rate given by $\dd\phi/\dd t=\Omega_{\mathrm{CR}}-\Omega$, tracing a line in the plots with slope $1/(\dd\phi/\dd t)$.

\begin{figure}
\includegraphics[width=\linewidth]{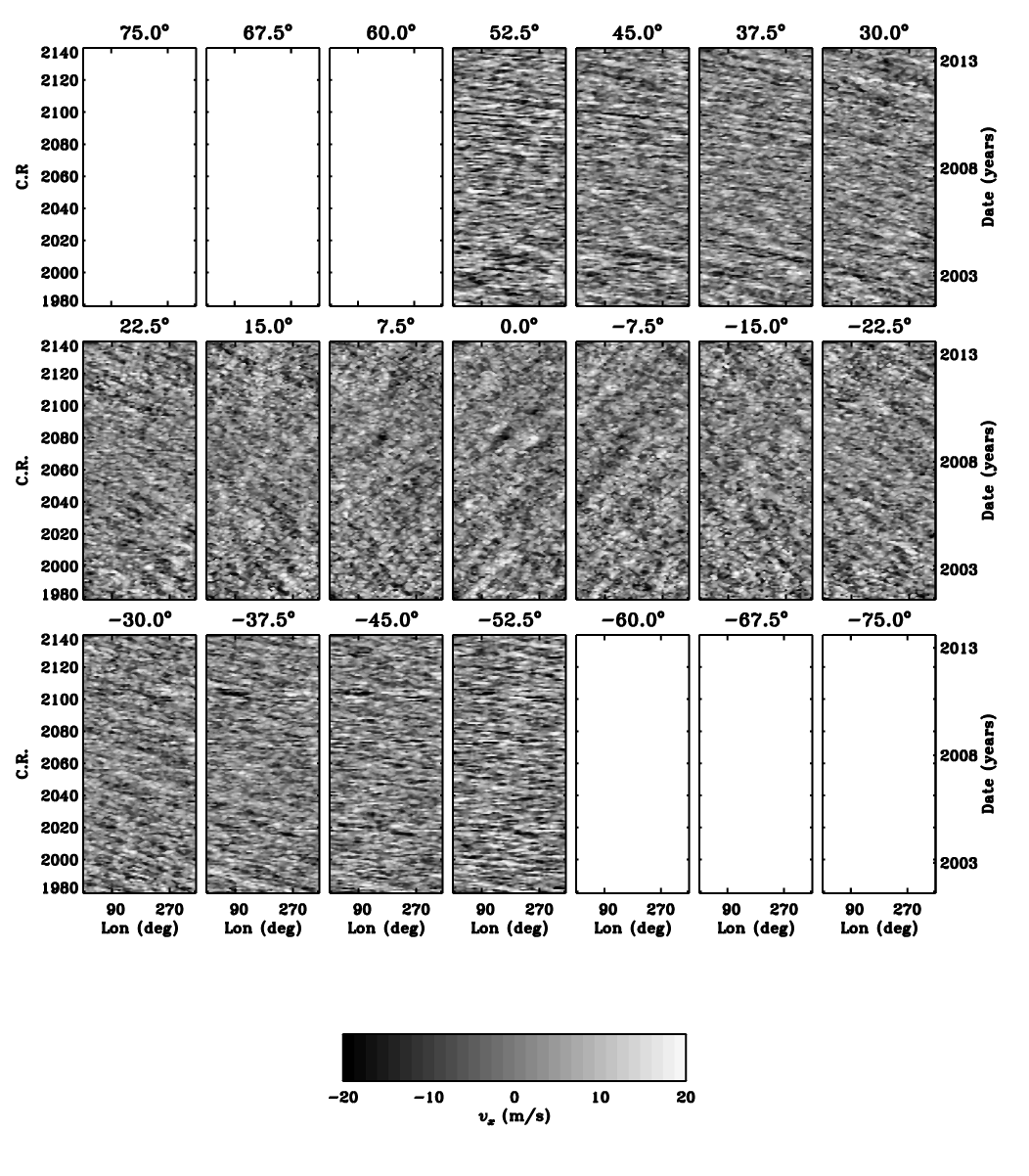}
\caption{
Longitude--time maps of zonal velocity residuals from GONG at each ring-diagram latitude. No data for GONG are available for latitudes poleward of $\pm 52.5^\circ$,  but the blank maps are shown to facilitate comparisons with later figures.}
\label{fig:fig1}
\end{figure}

\begin{figure}
\includegraphics[width=\linewidth]{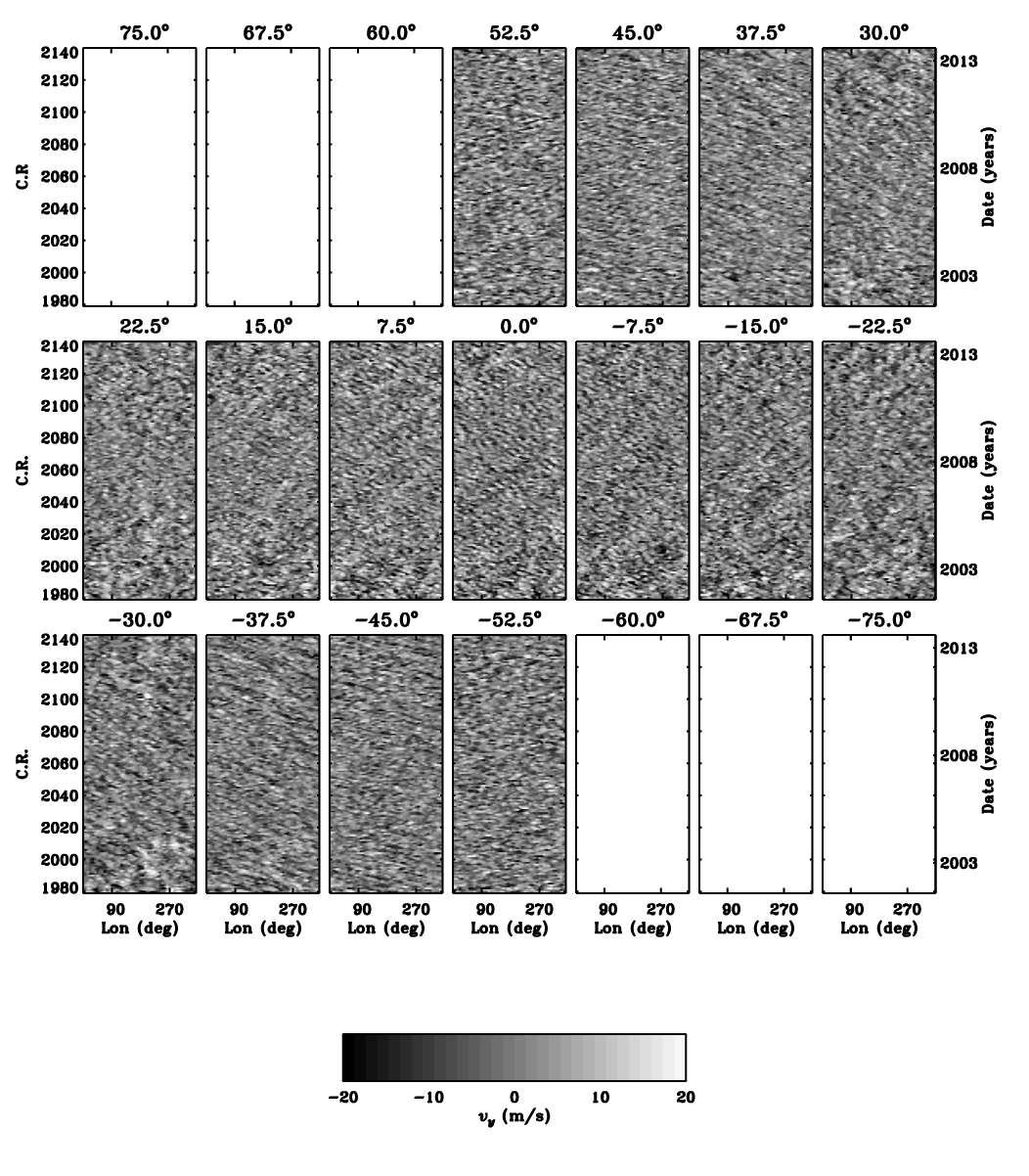}
\caption{
Longitude--time maps of meridional velocity residuals from GONG at each ring-diagram latitude. No data for GONG are available poleward of $\pm 52.5^\circ$,  but the blank maps are shown to facilitate comparisons with later figures.}
\label{fig:fig2}
\end{figure}

\begin{figure}
\includegraphics[width=\linewidth]{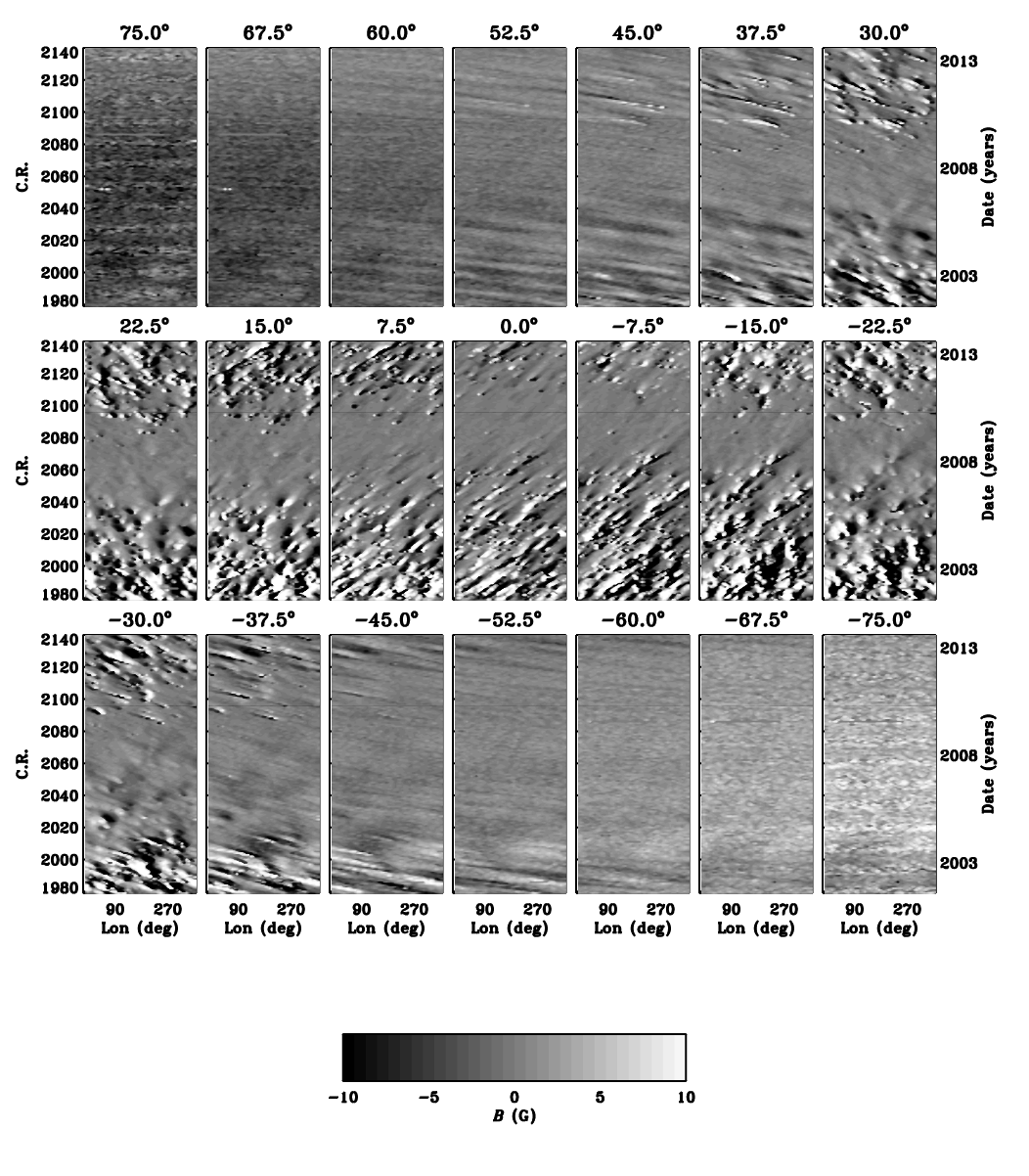}
\caption{
Longitude--time maps of MDI and HMI magnetic field strength at each ring-diagram latitude, averaged over the $15\times 15$$^{\circ}$ areas used for the ring-diagram analysis and plotted for the period covered by the GONG observations.}
\label{fig:fig3}
\end{figure}

Figure~\ref{fig:fig1} shows the zonal and Figure~\ref{fig:fig2} the meridional velocity residuals for the GONG ring data. Figure ~\ref{fig:fig3} shows the net magnetic field strength from MDI and HMI magnetograms, averaged over the same 15$^{\circ}$ patches and plotted in the same way as the flows. No temporal-average subtraction has been performed for the magnetic maps, and the highest latitudes clearly show the effects of $B_0$-angle variation, particularly for the MDI epoch. On this scale only strong isolated magnetic dipoles can be distinguished. The rise and fall of activity over the cycle is clearly visible in the magnetic maps; the amount of perturbation in the flows, as indicated by the standard deviation of the flow estimates at each rotation and latitude, also follows the activity cycle, but the flows show structure even in the quiet periods.  The zonal flow shows a slanted ``grain'' that appears to correspond to features moving in Carrington longitude due to the differential rotation, tracing upwards from left to right at low latitudes ($\dd\phi/\dd t > 0$, implying a rotation rate above the Carrington rate) and downwards at high latitudes (corresponding to rotation slower than the Carrington rate or $\dd\phi/\dd t < 0$). The further away the stripe is from the vertical, the larger the difference from the Carrington rotation rate. It is noticeable that while the $\pm 15^\circ$ latitudes show a downward slope in the velocity measurements the trend is upward in the magnetic data; this gives a hint that the two kinds of feature are revealing slightly different rotation profiles.

Some features persist for multiple rotations. For example, note the two ``bipole'' features around C.R. 2080 in the $0^\circ$ panel of Figure~\ref{fig:fig1}; each has a faster-than average flow on the lower-longitude side and slower on the higher-longitude side, and persists for a few rotations. These features do not appear to correspond to any magnetic feature that is pronounced enough to show up on the magnetic plots, but the configuration of the flows suggests an inflow into a ``sink'' feature. Another feature is the broad, low-contrast, but prolonged streak of below-average velocity starting at the far left of the $7.5^\circ$ South panel around C.R. 2040 and extending over at least 40 rotations. Both of these features appear during the period of extremely low solar activity that marked the Cycle 23\,--\,24 solar minimum, as can be seen by comparing with Figure~\ref{fig:fig3}. At the higher latitudes, for example $45^\circ$ North, we see shallowly downward-slanting stripes that tend to have below-average speed on the lower-longitude side, which would correspond to flows outward from a persistent ``source'' feature. These patterns are more noticeable during the higher-activity periods and less visible during solar minimum, which supports the conjecture that they are associated with areas of activity.

The meridional flows (Figure~\ref{fig:fig2}), on the other hand, show very little trace of any persistent features. This is what we would expect if the bipolar features arise from flows into or out of features such as regions of diffuse surface magnetic flux; the poleward flows would lie to the north and south of the feature at the same longitude and would appear in different latitude ``bins''  (and hence in different panels of the plot) or would tend to cancel one another if they fall within the same bin; the features would therefore be less visible. For flows circulating around a feature, on the other hand, we would expect a stronger signature in the meridional-flow maps where the poleward flows would appear to the east and west of the feature and flows parallel to the Equator to the north and south.

\begin{figure}
\includegraphics[width=\linewidth]{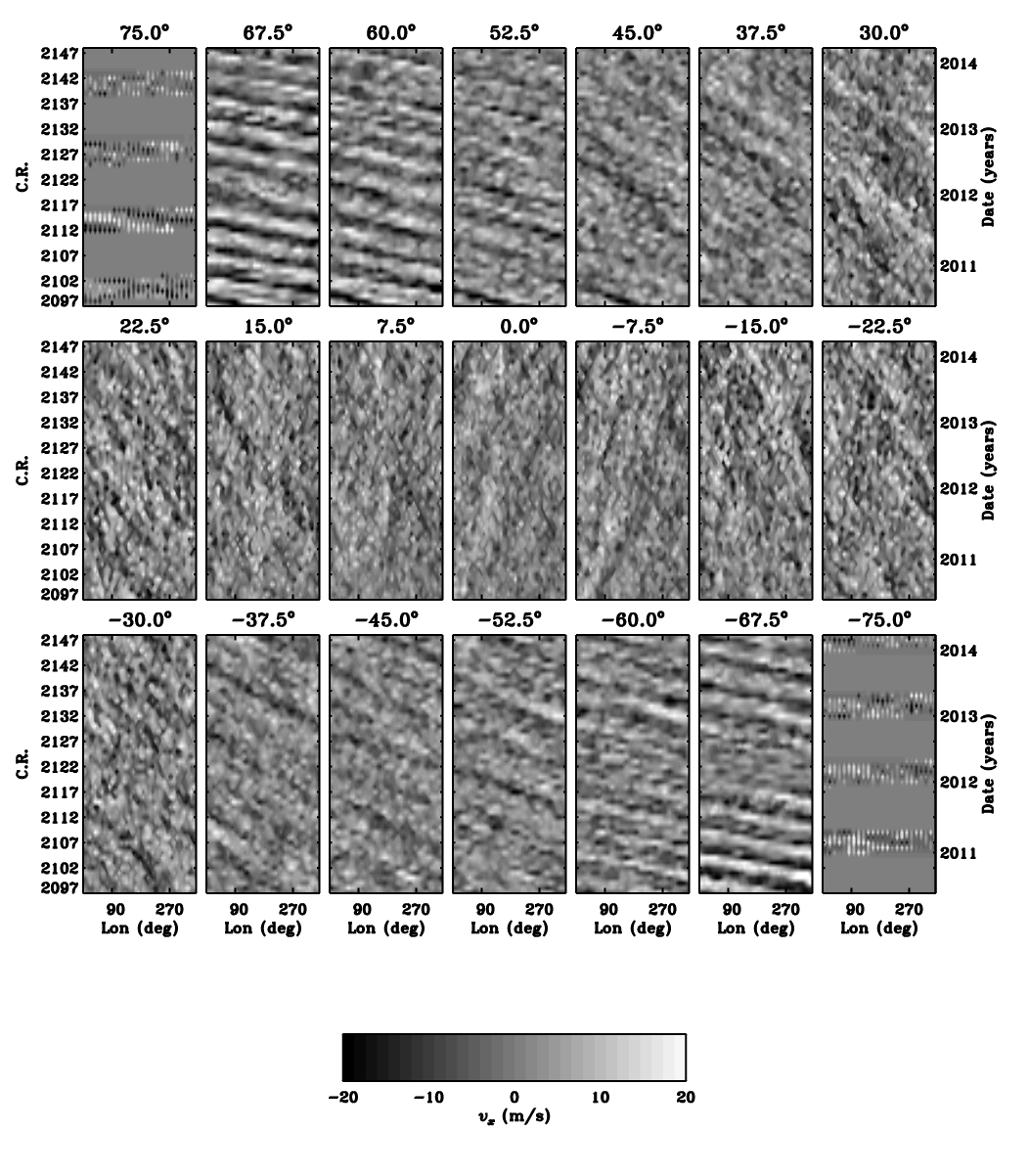}
\caption{
Longitude--time maps of zonal velocity from HMI at each ring-diagram latitude.} 
\label{fig:fig4}
\end{figure}

\begin{figure}
\includegraphics[width=\linewidth]{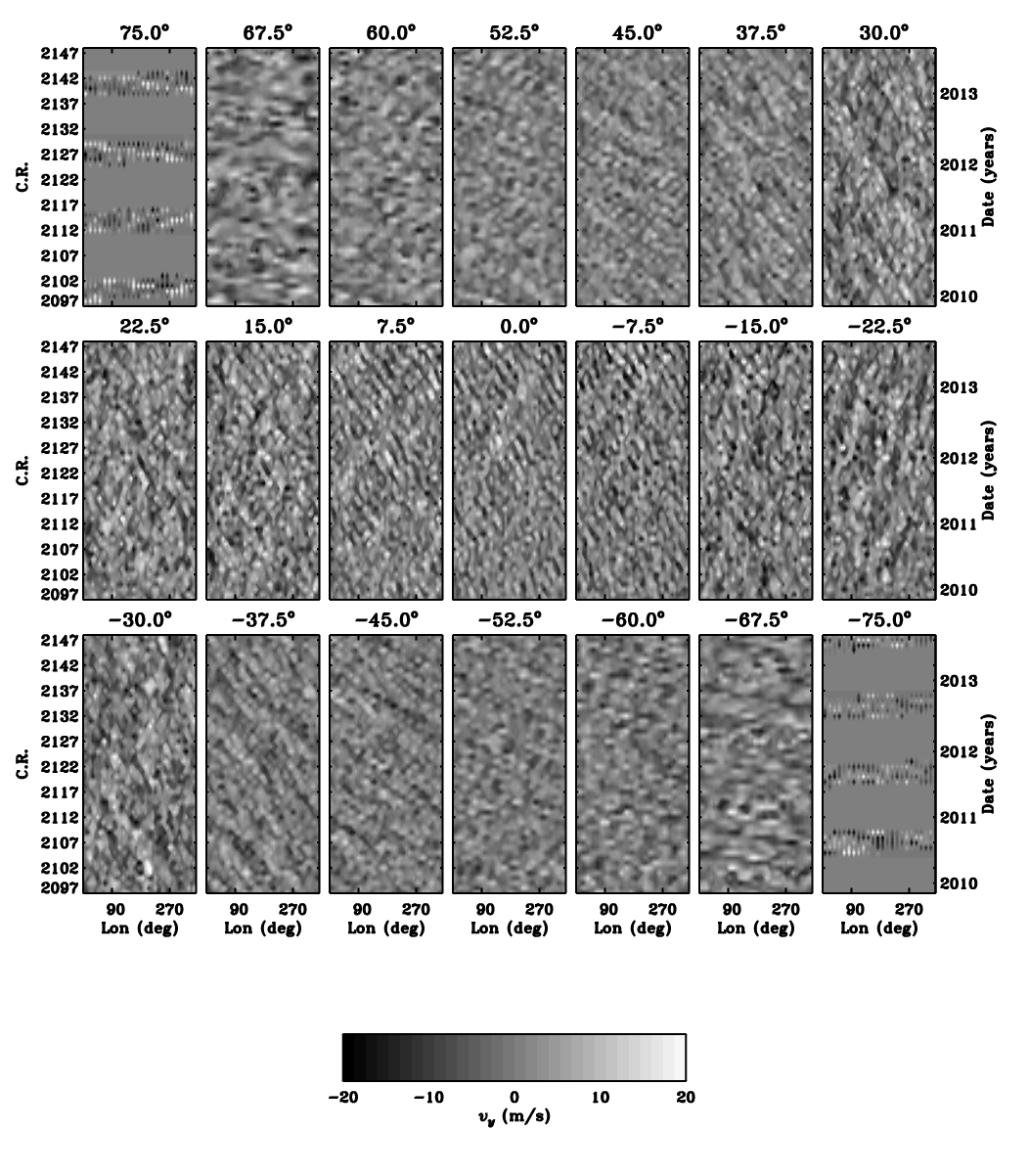}
\caption{Longitude--time maps of meridional velocity from HMI at each ring-diagram latitude.}
\label{fig:fig5}
\end{figure}

Figures~\ref{fig:fig4} and~\ref{fig:fig5} show a similar analysis of the ring-diagram flow data from HMI. 
The lower noise and higher resolution of the HMI data reveal the streak pattern more clearly; furthermore, at the higher latitudes that are not covered by GONG we see a very clear pattern of slanted stripes of slower and faster flows, and there is also a detectable slanted grain in the meridional flows, particularly at the higher latitudes where we do see a few bipolar features. The r.m.s. value for the zonal-flow variations at $67.5^\circ$ is about 10\,m\,s$^{-1}$. This amounts to about 1\,--\,2\,\% of the near-surface rotation rate at that latitude; it is also a few times larger than the solar-cycle variation of the rotation rate due to the poleward branch of the torsional oscillation \citep{2001ApJ...559L..67A,2013ApJ...767L..20H} and comparable with the amplitude of the large-scale meridional flow.

\begin{figure}
\includegraphics[width=\linewidth]{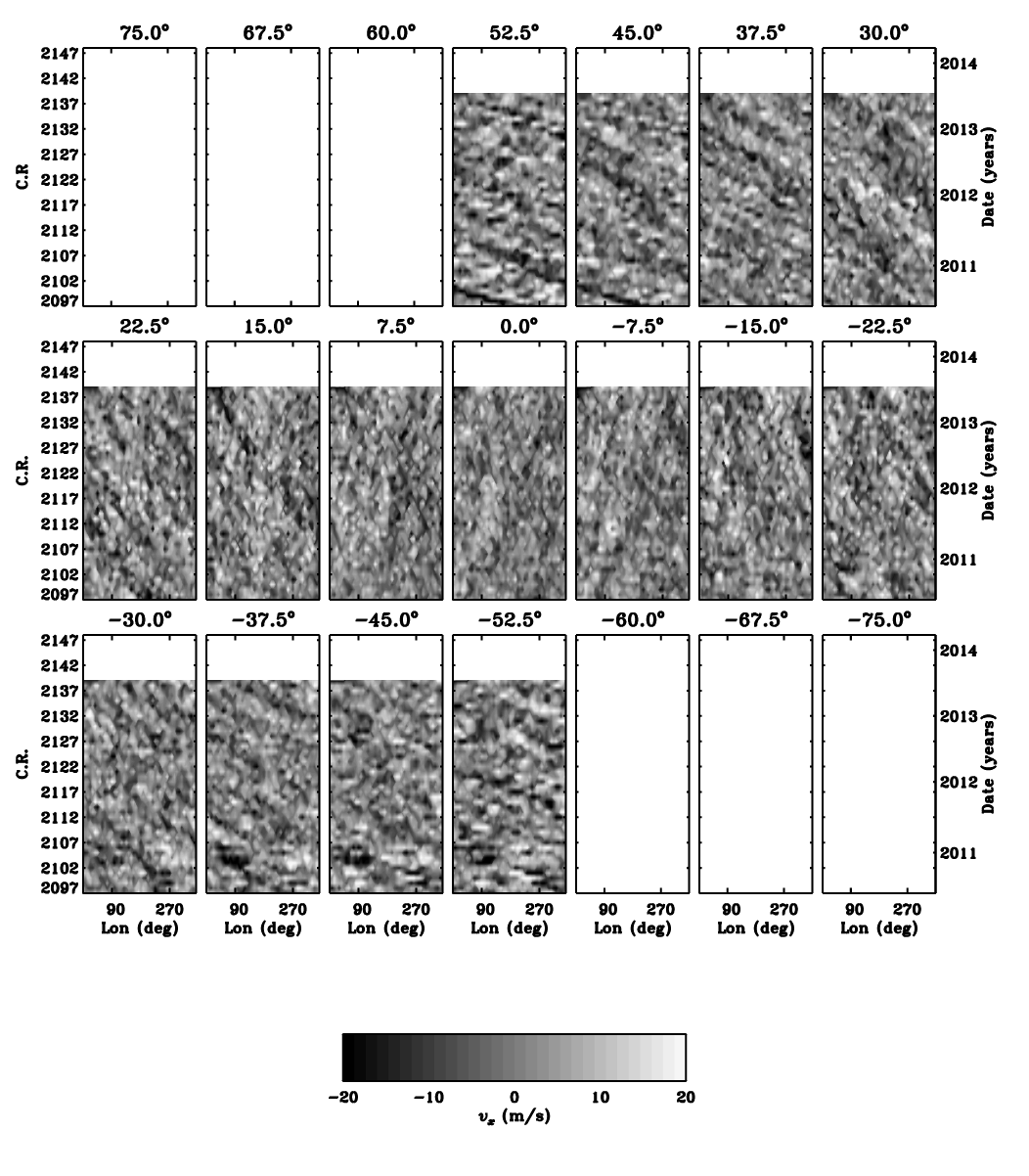}
\caption{Longitude--time maps of zonal velocity at each ring-diagram latitude during the period of the HMI observations, from GONG data.
}
\label{fig:fig6}
\end{figure}

For comparison purposes, in Figure~\ref{fig:fig6} we replot the GONG zonal flows on the same scale as Figure~\ref{fig:fig4}. Over the 15 latitudes and 43 rotations common to both data sets the correlation coefficient between GONG and HMI is 0.81 for the zonal and 0.84 for the meridional velocity. This gives us some confidence that the flow patterns are neither instrumental in origin nor pure noise.

\begin{figure}
\includegraphics[width=\linewidth]{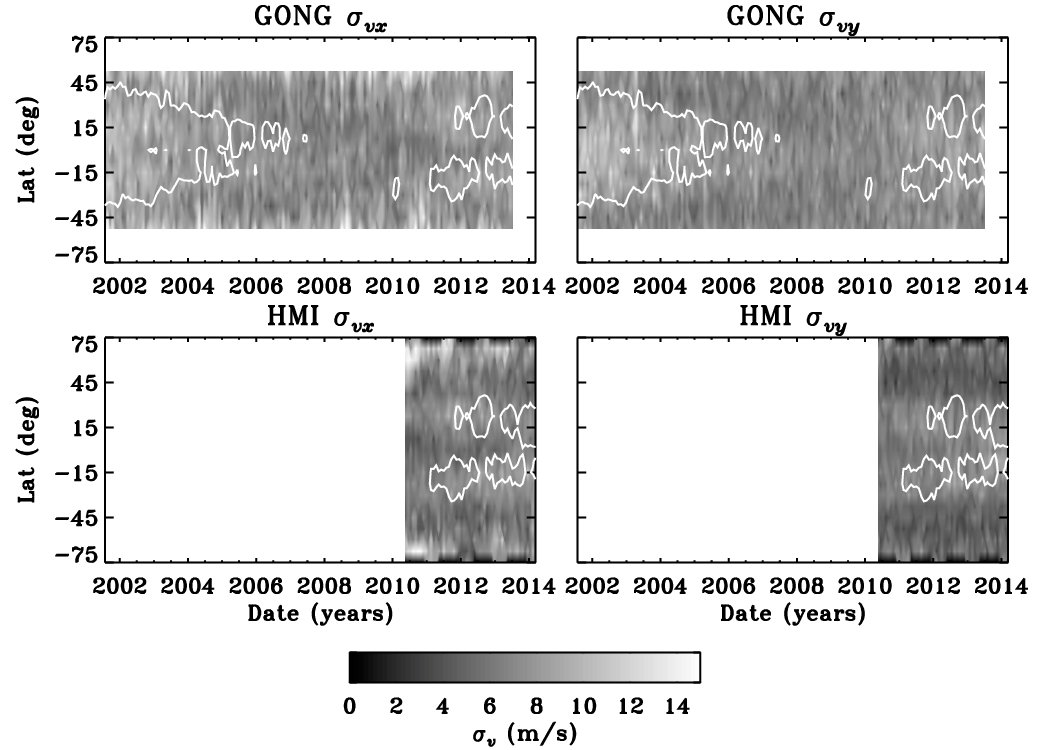}
\caption{Time--latitude maps of the standard deviation of the zonal (left) and meridional (right) flow residuals for GONG (top) and HMI (bottom). The overlaid contours represent the 5G level of the unsigned magnetic field strength.}
\label{fig:figt}
\end{figure}

In Figure~\ref{fig:figt} we show the standard deviations of the zonal and meridional flow residuals within each rotation and latitude bin, overlaid with the 5 gauss contour of the unsigned magnetic field strength. These plots clearly show that the mid-latitude flow residuals -- both zonal and meridional -- tend to have more structure in the presence of magnetic activity, but the correlation is by no means perfect.

 Curiously, the zonal flow residuals in the $+60^\circ$ band have a correlation coefficient of -0.41 with those in the $-67.5^\circ$ band, while the correlation coefficient is $-0.27$ for the $\pm 60^\circ$ bands and negigible elsewhere. For the meridional flow residuals the only somwhat significant correlation between opposite latitudes is $+0.33$ between the bands at $\pm 7.5^\circ$. There is also a correlation coefficient of $+0.30$ between the $\pm 7.5^\circ$ latitudes in the GONG meridional flow. Could these correlations hint at, on the one hand, a global-scale variation of the high-latitude zonal flow on a timescale of months, and on the other, low-latitude meridionally-aligned underlying features reminiscent of the ``banana cells'' seen in convection simulations?


\begin{figure}
\includegraphics[width=\linewidth]{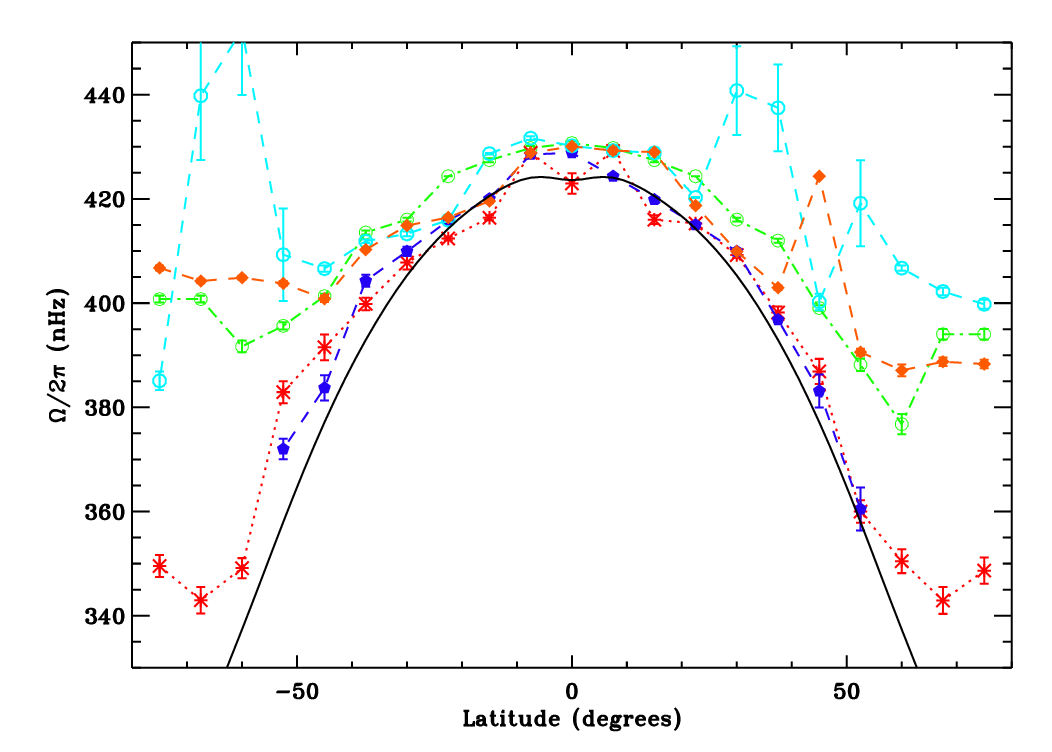}
\caption{
Synodic rotation rate as a function of latitude from global inversions at $0.99\,{\mathrm{R}}_{\odot}$(solid curve), compared with the rate inferred from cross-correlation analysis of HMI synoptic magnetic data (green open circles, dot-dashed line), EIT {195 \AA} synoptic charts (cyan open circles, dashed line), GONG PFSS coronal hole proxy (orange filled squares, dashed line), and zonal flow data from GONG (blue filled circles, dashed line) and HMI (red stars, dotted line) ring diagrams. 
}
\label{fig:fig7}
\end{figure}





\subsection{Cross-Correlation Analysis}
The features that we see as patterns in the residual flow field appear to persist over multiple rotations; in what follows, we assume that they have some identity -- perhaps tied to magnetic features not obvious in synoptic maps -- that can withstand the advection and shear of the wider flow field. Meridional advection might be expected eventually to carry such an entity out of its original 15-degree latitude band; the band covers about 180\,Mm in horizontal distance at the surface, so the time for this to occur at a meridional flow rate of $10\,\mathrm {m\,s^{-1}}$ would be about 100 days, or roughly four rotations, on average, although it might be longer for a feature that started well to the equatorward side of the latitude range. There will also be shearing effects; the shear from differential rotation ranges from zero at the Equator to about  $75\,\mathrm{m\,s^{-1}}$ per degree at $45^\circ$ latitude; a feature in the band centred on $7.5^\circ$ latitude and spread over two degrees of latitude would experience a shear of  $\approx 20\,\mathrm{m\,s^{-1}}$ and so might be expected to be completely destroyed in around two rotations if nothing was acting to maintain its identity, but we know that active regions can persist for at least that long. In what follows, we will treat the features as entities that are carried along -- like eddies in a stream -- by the differential rotation but are not destroyed by it.

In order to verify that the migration of the features we see corresponds to the differential rotation, we carry out a cross-correlation analysis on the longitude--time plots. For each slice at longitude [$\phi$] through a map at a given latitude [$\theta$] we form the temporal cross-correlation function [$C_\phi(v(\phi,T),v(\phi+\delta\phi,\delta T))$] of $v(\phi,T)$ with $v(\phi+\delta\phi,T)$ at longitude offsets out to $\delta\phi=\pm 17\Delta\phi$, where $\Delta\phi$ is the longitude step (in degrees) and $\delta T$ is the temporal lag (in Carrington rotations). This number of steps was empirically chosen to allow a reasonable amount of averaging while avoiding ``wrap-around'' effects. The cross-correlation maps for each latitude are then averaged across all longitudes [$\phi$] to give a two-dimensional mean correlation map [$C_{\rm av}(\delta\phi,\delta T)$].  The maximum of the function is then taken at each time offset and a trend with $\delta T$ fitted to the maximum-correlation longitude offset [$\delta\phi_{\mathrm{max}}(\delta T)$]. In this fit, zero-offset values are omitted as there is some distortion of the profile due to end effects. This fit yields a differential rotation rate [$\dd\delta\phi/\dd \delta T$] in units of degrees per Carrington rotation at each latitude, which can then be converted to a synodic angular rotation rate in nanohertz. In Figure ~\ref{fig:fig7} we show the rotation-rate estimates from GONG and HMI ring-diagram zonal flows compared with those from HMI magnetic data over the period (C.R. 2097\,--\,2048), or May 2010 to March 2014, corresponding to the HMI observations. We show on the same axis the rotation rate from global analysis of GONG data over the period from mid-2001 to mid-2013 corresponding to the GONG ring-diagram data. The error bars shown on these plots are probably underestimated; however, we can see that the rotation rate from magnetic features is systematically slightly higher than those from the global helioseismic analysis and the flow correlations. 

Also included in Figure~\ref{fig:fig7} are results from the cross-correlation analysis of two datasets related to the corona: the {\em Extreme ultraviolet Imaging Telescope} (EIT) {195\,\AA} intensity and the potential field source surface predictions of coronal hole locations based on GONG magnetograms. While the coronal rotation estimates are quite close to the magnetic one at lower latitudes (where the analysis will be dominated by emission associated with active regions), at higher latitudes it is faster. 
In the EIT synoptic maps, as reflected by the cross-correlated rotation curve, different features -- dark coronal holes closer to the poles and bright active regions nearer the Equator -- dominate the appearance of the corona at different latitudes.
The coronal hole boundaries are not tied to the same magnetic features, but there is a systematic (westward) displacement, which appears to result in an increased rotation rate at latitudes higher than 50$^{\circ}$. At lower latitudes, {\it i.e.} in the activity belt, bright coronal features that are genuinely tied to magnetic features are dominant (active regions), and therefore the EIT intensity rotation curve matches that of the magnetic features.

Another very interesting result shown in Figure~\ref{fig:fig7} is the higher ({\em i.e.} less differential) rotation rate profile of the HMI magnetic features with latitude. As most of the high-latitude magnetic features originate from the activity belt, one may perhaps speculate that as they random-walk and spread towards higher latitudes, they may retain some of their original angular momentum. Another possibility is that they maintain some connections to their faster-rotating roots and Lorenz forces keep their rotation rate less differential.

\section{Discussion and Conclusions}

We have found that some near-surface horizontal flow patterns can persist over multiple rotations, even in relatively quiet Sun. The migration of these features in Carrington longitude, as determined by a cross-correlation analysis of the longitude--time maps,  corresponds to a rotation rate that is close to the near-surface rotation rate from global helioseismology and slightly lower than the rate found from a similar analysis of magnetic data. Given that the near-surface rotation rate is known to increase with depth \citep{2002SoPh..205..211C}, this suggests that the flow patterns may be less deeply ``rooted'' in the outer convection zone than the magnetic structures. This is similar to, for example, the result of \inlinecite{1999SoPh..184...41J}, who found that sunspots rotate at a faster rate than that determined from surface Doppler velocity measurements.
The existence of persistent flow patterns associated with diffuse magnetic flux after the decay of active regions may point to an explanation for the 
presence of modulations (the ``torsional oscillation'') in the large-scale flow that follow the solar cycle but are evident even at solar minimum.

At high latitudes, we see strong (10 m\,s$^{-1}$) variation in the zonal flow structure that appears to represent a pattern of regions of faster and slower zonal flow rotating at or above the helioseismic near-surface rotation rate. These patterns do not appear to be directly related to high-latitude coronal features, as they correspond to a rotation rate lower than that found for coronal holes at the same latitude.
\inlinecite{2013Sci...342.1217H} detected large-scale high-latitude flow structures in supergranular data from the first few Carrington rotations of HMI data and attributed them to giant convection cells. We believe that in our high-latitude ``stripe'' pattern we are seeing the same phenomenon of extended zonal-flow structures. The anticorrelation between northern and southern hemisphere results is suggestive of an ``oscillation'' at global scale and on a timescale of months. We also see a possible hint, in the positive correlation of meridional flow patterns across the equator, of large-scale flows aligned parallel to the rotation axis.

It will be interesting to see what happens to these patterns as the solar cycle declines towards the next minimum.

%
\begin{acks}
We acknowledge the Leverhulme Trust for funding the ``Probing the Sun: inside and out'' project upon which this research is based. RH was partially supported by the BiSON group, which is funded by the Science and Technology Facilities Council (STFC), and also acknowledges computing support from the National Solar Observatory. LvDG acknowledges the Hungarian government for grants OTKA K-109276. DB thanks STFC for support under Consolidated Grant ST/H00260/1. RK was supported by NASA grant NNX11AQ57G to the
National Solar Observatory.  This work utilizes data obtained by the Global Oscillation Network
Group (GONG) program, managed by the National Solar Observatory, which
is operated by AURA, Inc. under a cooperative agreement with the
National Science Foundation. The data were acquired by instruments
operated by the Big Bear Solar Observatory, High Altitude Observatory,
Learmonth Solar Observatory, Udaipur Solar Observatory, Instituto de
Astrof\'{\i}sica de Canarias, and Cerro Tololo Interamerican
Observatory.
HMI data courtesy of NASA/SDO and the HMI science team.

\end{acks}

%
%
 \bibliographystyle{spr-mp-sola}
 \bibliography{ms}  
%
%
%
%

\end{article} 
\end{document}